\documentclass[prb,aps, babel,twocolumn]{revtex4}

\usepackage{amsmath,amssymb,amstext,amscd,amsthm,amsopn,graphicx,indentfirst,mathrsfs}
\usepackage{epstopdf}

\newcommand{\w}{\omega}
\newcommand{\im}{\Im\mathfrak{m}}
\newcommand{\R}{\textsc{r}}
\renewcommand{\vec}[1]{{\bf #1}}
\renewcommand{\L}{\textsc{l}}

\renewcommand{\t}{\textsc{t}}

\newcommand{\e}{\textsc{e}}
\newcommand{\vm}[1]{\big\langle #1 \big\rangle}
\newcommand{\Cal}[1]{\mathcal{#1}}

\newcommand{\p}{\partial}
\newcommand{\g}[1]{{\bf #1}}
\newcommand{\prt}[1]{\left( #1 \right)}
\newcommand{\XXX}[1]{}

\begin{document}
\title{Heat transport of clean spin-ladders coupled to phonons:\\
Umklapp scattering and drag}

\author{E. Boulat$^{1,2}$, P. Mehta$^2$, N. Andrei$^2$, E. Shimshoni$^3$, A. Rosch$^4$}

\affiliation{
$^1$Laboratoire MPQ, CNRS, Universit\'e Paris Diderot, 75205 Paris Cedex 13, France\\
$^2$Center for Material Theory, Rutgers University, Piscataway, NJ 08854, USA\\
$^3$Department of Math-Physics, University of Haifa at Oranim, Tivon 36006, Israel\\
$^4$Institute for Theoretical Physics, University of Cologne, 50937 Cologne, Germany}

\begin{abstract}
We study the low-temperature heat transport in clean two-leg spin
ladder compounds coupled to three-dimensional phonons. We argue that
the very large heat conductivities observed in such systems can be
traced back to the existence of approximate symmetries and
corresponding weakly violated conservation laws of the effective
(gapful) low--energy model, namely pseudo-momenta. Depending on the
ratios of spin gaps and Debye energy and on the temperature, the
magnetic contribution to the heat conductivity ($\kappa_{mag}$) can
be positive or negative, and exhibit an activated or anti-activated behavior.
In most regimes $\kappa_{mag}$ is dominated
by the spin-phonon drag: the excitations of the two subsystems
have almost the same drift velocity, and this allows for an estimate of the ratio $\kappa_{mag}/\kappa_{ph}$ of the magnetic and phononic contributions to the heat conductivity.
\end{abstract}
\maketitle

\section{Introduction}

Recent experiments both on spin-chain \cite{chain} and ladder
compounds \cite{ladder} showed a surprisingly large magnetic
contribution to the heat conductivity: the heat conductivity in the
direction parallel to the ladder (attributed to magnons \emph{and}
phonons) largely exceeds the heat conductivity in directions
perpendicular to it (attributed to the phonons alone). The heat
conductivity of clean \emph{gapless} spin $1/2$ chains coupled to
phonons shows a simple exponential behavior
 \cite{shimshoni03}, associated with a single characteristic energy
 scale resulting from the high-energy process needed to relax momentum.
In contrast, gaps open up in the spectrum of (two-leg) spin ladders,
and consequently the heat transport  involves a complex interplay
between different energy scales leading to a rich gamut of possible
behaviors. In this paper, we present a theoretical framework to
describe low-temperature heat transport in such clean  \emph{gapped}
quasi one-dimensional systems when they are coupled to phonons.

In the absence of disorder, heat transport in quasi one-dimensional
systems is determined by momentum conservation (or more precisely by
`pseudo-momentum' conservation \cite{rosch00,rosch01,shimshoni03}).
In a clean lattice, momentum transfers are quantized and therefore
the momentum can only decay via an Umklapp process involving a
large-momentum high-energy state. This implies that transport in
such systems is non-universal as it depends on both high- and
low-energy features. Therefore controlled analytic calculations are
usually not possible (the situation is, however, simpler for systems
with a finite magnetization \cite{rosch00,shimshoni03}).
Nevertheless, we shall show that under certain circumstances such
calculations are possible.
Indeed, in many spin-systems, the typical spin-velocity, $v_s$, is large
compared to the sound velocity of the acoustic phonons, $v_p$.
Therefore, the large momentum state with the lowest energy (required
for an Umklapp process) will have most of its momentum carried by
phonons. This has two consequences: heat transport is (i) dominated
by spin-phonon scattering and is (ii) determined by high-energy
features of the phonon system but low-energy properties of the spin
system. The latter observation implies that controlled calculations
of transport are in principle possible (up to non-universal
  prefactors describing the electron-phonon coupling) using the fact that
high-energy properties of the weakly interacting phonons are often
known or can be measured. The necessary low-energy correlators of
the gapped spin-system can be obtained from an effective field
theory which can be analyzed by semi-classical \cite{sachdev97b}
or form-factor \cite{altshuler05} methods for temperatures $T$
below the gap.

In contrast, in pure systems where phonons are absent (e.g. cold
atom realizations), or when $v_p>v_s$, no such controlled
calculation is possible as little is known about the non-universal
high-energy properties of strongly interacting spin systems and one
can make only qualitative statements, e.g. that the heat
conductivity is exponentially large but finite \cite{shimshoni03}.
Previous numerical studies of pure ladder systems at high $T$
indicate that the heat conductivity of spin ladders is finite
\cite{zotos03a,heidrich03} but could not reach temperature of the
order of the gap and below. On the analytical side, few results are
available. The field theoretical treatment of Ref.[\onlinecite{orignac03}]
ignored the role of Umklapp processes, thus leading to ballistic
transport in a clean and pure system. Rozhkov and Chernyshev
\cite{chernyshev05} included the effect of disorder and phonons in
spin chains within a Boltzmann equation approach, but did neither
consider spin-phonon drag nor  addressed the question of Umklapp,
which becomes essential in a clean system.

In the following, we will first present the general field theoretical
framework on which our calculation is based. We discuss the various
Umklapp processes that induce (pseudo-) momentum decay in quanta of
size $G$ or $G/2$, where $G$ is the reciprocal lattice vector. We
shall discuss the role of such processes in (weakly) violating the
conservation laws of the model, allowing a hydrodynamic description
of the system. We then use the memory matrix formalism to calculate
the heat conductivity $\kappa$ in the various regimes.
The memory matrix
approach is generally {\em not} exact even in the limit of small
couplings but it
can be shown \cite{jung07} to give a
lower bound to $\kappa$, which is saturated in
the limit of a large separation of time scales between slow and fast
modes, i.e. if a hydrodynamic description is possible, as is the case
here. Finally, we interpret our results in terms of the spin-phonon
drag and the decay rates of various slow modes.
The various resulting behaviors of $\kappa(T)$ are summarized in Fig. 1,
where regime I is consistent with the
particular data of Ref.[\onlinecite{ladder}] for $T$ below the spin gap, but still larger than a lower scale $T_*$, where the magnetic heat conductivity displays an activated behavior.

\section{low energy effective theory}

Our starting point is the following Hamiltonian
\begin{equation}
H=\sum_\gamma H_s(\gamma) + H_p + H_{s,p}
\end{equation}
which describes an array of spin-ladders (denoted by `$s$',
with $\gamma$ the ladder index)  coupled
to acoustic three-dimensional phonons (`$p$'),  via the term
$H_{s,p}$.
For simplicity, we assume $H_p\approx\sum_{{\bf k},\alpha} v_\alpha
|{\bf k}|  a^\dagger_{{\bf k},\alpha} a_{{\bf k},\alpha}^{}$ where
the velocities of the various branches of acoustic phonons
($v_\alpha$) are approximated by a characteristic velocity $v_p$,
associated with the Debye energy via $\Theta_D \sim v_p G/2$.

A single spin ladder is described by
\begin{equation}
 H_s=J_\parallel\sum_{j,\ell}\g{S}_{\ell,j}\cdot\g{S}_{\ell,j+1}
 + J_\perp\sum_{j}\g{S}_{1,j}\cdot\g{S}_{2,j},
 \label{HamMin}
\end{equation}
where $\g{S}_{\ell,j}$ is a spin-$\frac{1}{2}$ operator acting on site $j$ and on leg $\ell=1,2$ of the ladder.

As we are interested in the heat conductivity at low $T$, it is
useful to consider the effective low-energy theory for $H_s$
(assuming a small gap, $\Delta \ll J_\parallel$) described in terms
of four massive Majorana fermions \cite{shelton96}
\begin{eqnarray}
 H_{s} &=& H_0^s
 + \sum_i g_i \int \,d x \; {\cal
    O}_i(x)\label{HeffThermo}   \ \\
 H_0^s &=& \int \! d x \sum_{a=0}^3  \frac{i v_a}{2}\left(
 \xi^a_\L\p_x\xi^a_\L - \xi^a_\R\p_x\xi^a_\R\right) +i
 (-)^{\delta_{a0}}\Delta_a \; \xi^a_\R\xi^a_\L\; .\nonumber
\end{eqnarray}
In the above expression, the operators ${\cal O}_i$ are all
irrelevant and marginal operators allowed by the symmetry of the
original lattice Hamiltonian (\ref{HamMin}). The three Majorana
fields $\xi^a$, $a=1,2,3$, describe the low lying magnon triplet
with the velocity $v_a=v_1$ and (spin) gap $\Delta_a=\Delta_1$, while the
remaining Majorana field, $\xi^0$, describes a singlet excitation
with gap $\Delta_0$, and velocity $v_0$: the single-particle excitations have dispersion relation $\epsilon_a(k)=\sqrt{v_a^2 k^2+\Delta_a^2}$. While $\Delta_0/\Delta_1\sim 3$  for weak
$J_\perp$, this ratio changes for larger
$J_\perp/J_\parallel$  or when other microscopic interactions are
present -- so that we shall consider it as a free parameter,
yet assuming $\Delta_0>\Delta_1$.

 The total heat current, obtained through
the continuity equation of the energy density, is
\begin{eqnarray}
 J_\e \approx v_p^2 P_p+v_0^2 P_0+v_1^2  P_1
 \label{Je}
\end{eqnarray}
where $P_p$, $P_0$ and $P_1$ are the momentum operators of
phonons, singlets and triplets, respectively, for example $P_1=
-\frac{i}{2}\sum_{a=1...3,\gamma} \int dx\, \prt{\xi^a_{\gamma\R}
\p_x\xi^a_{\gamma\R} + \xi^a_{\gamma\L}\p_x\xi^a_{\gamma\L}}$. In
Eq.~(\ref{Je}) we neglect further contributions from the
interactions which
turn out to give only subleading corrections.

The crucial observation on which our following analysis is based,
is that the effective low-energy theory of our initial Hamiltonian
conserves the total momentum,
\begin{equation}
 P_\t=P_p+P_0+P_1.
 \label{Pt}
\end{equation}
A direct consequence is that -- within this low-energy description
-- the heat conductivity is infinite \cite{mazur69,zotos03b}.
However, the continuous translational invariance of
(\ref{HeffThermo}) is not a true symmetry and follows from
neglecting Umklapp terms whose inclusion leads to a decay of $P_\t$,
which now acquires an exponentially long life-time at low
temperature. This is to be contrasted to all other decaying modes,
whose life-time behave as power laws of $T$, and allows for a
hydrodynamic description based on the slow modes $P_a$, $a=p,0,1$.
Including these Umklapp operators is thus the correct way to obtain
a low-energy effective theory suited to the calculation of transport
properties.

Three different Umklapp terms turn out to be important if we
assume that $v_p<v_0,v_1$
\footnote{Additional Umklapp terms, acting only in the spin sector, involve much higher energy states in the limit of soft phonons, and thus have exponentially suppressed effects}:
\begin{eqnarray}
 H^p_U &=& \sum_{n\geq 3}g_p^{(n)} \int dx \,(\p_x q)^n
 \cos(G x)
 \label{pUmk} \\
 H^{sp}_U &=& g_D^{sp}\int dx\, (\p_x q)^2\Cal O_D(x) \cos(G x/2)
 \label{spUmk1} \\
 &+& i\sum_a g^{sp}_a \int dx\, (\p_x q)^2\xi^a_\R\xi_\L^a
 \cos(G x)
 \label{spUmk}
\end{eqnarray}
with $q(x)\propto \int \frac{1}{\sqrt{v_p |\vec{k}|}} e^{i k_x x}
(a^\dagger_{\vec{k}}+a_{-\vec{k}}^{}) \frac{d^3 \vec{k}}{(2 \pi)^3}$
the displacement field for acoustic phonons projected along the
ladders ($a^\dagger_{\vec{k}}$ is an abbreviation for the sum
of contributions from various phonon modes).
$\Cal O_D \equiv \prod_{a=0}^3 \sigma^a$ is the
continuum limit of the dimerization operator
 $(-)^j \big(\vec S_{1,j}\!+\!\vec S_{2,j}\big)\cdot
\big(\vec S_{1,j+1}\!+\!\vec S_{2,j+1}\big)$,
where each $\sigma^a$ $(a=0..3)$ is the Ising spin operators for
the quantum Ising model that is naturally associated to
 the Majorana $\xi^a$ theory \cite{shelton96} (in our
conventions, Ising model '0' is in its quantum disordered phase
while the three other models are ordered).
We checked that linear couplings to  phonons in (\ref{spUmk}) are
subdominant
\footnote{Scattering events preserving momentum and energy when a single phonon is involved, with momentum transfer of order $G$, force at least one magnon to carry momentum of order $G$, and thus involves higher energy states.}.

On top of the Umklapp terms it is also important to include normal processes
\begin{equation}
 H_{N,k}^{sp}=i\sum_a g^{N,k}_a \int dx\, \left(\p_x
 q\right)^k\,\xi^a_\R\xi_\L^a
 \label{normalsp-ph},
\end{equation}
which allow momentum exchange between the spin and phonon systems.
We do not consider normal operators acting only in the spin sector
or only in the phononic sector, as they commute \emph{separately}
with the spin and phonon momentum operators, and accordingly do not
contribute to leading order to the heat conductivity.

\section{hydrodynamic approach}

To obtain a hydrodynamic description we first identify a basis in
the space of slow modes, in our case given by $P_p, P_0$ and $P_1$.
Then we introduce the matrix of static susceptibilities of the slow
modes, $\chi_{ij}=\frac{1}{LT}\vm{P_i(0) P_j(0)}$ (equal time
correlator), and a matrix of conductivities defined by Kubo formulas
for the $P_i$'s, $\sigma_{ij} = \frac{T}{\omega}\int_0^{1/T}
d\tau\,e^{i\omega\tau}\vm{\mbox{T}_\tau P_i(\tau)P_j(0)}$,
$i,j=p,0,1$. The heat conductivity is then given by
\begin{equation}
\kappa=\frac{1}{T} \sum_{i,j=p,0,1} v_i^2 v_j^2 \, \sigma_{ij}
\equiv \kappa_{ss}+ 2\kappa_{sp}+ \kappa_{pp}\; .
\label{kappasum}
\end{equation}
Note that besides the spin and phonon heat conductivity $\kappa_{ss},
\kappa_{pp}$ there is also the drag term $2\kappa_{ps}=\frac{2}{T}
\sum_{b=0,1} v_p^2 v_b^2 \sigma_{b p}$.

Within the memory-matrix approach \cite{zwanzig61,giamarchi}, the
matrix of conductivities $\hat{\sigma}(\w)$ is expressed as
\begin{equation}
 \hat{\sigma}(\omega,T)=\hat\chi\,\frac{1}{\hat
  M(\w)-i\omega\hat\chi}\,\hat\chi
 \label{Sigmamatform}
\end{equation}
where $\hat {M}(\w)$ is the so-called memory matrix. It can be shown
that to leading order in the coupling constants of the Umklapp
terms, $g^U$, and of the normal spin-phonon term, $g^N$, the memory
matrix is simply given by \cite{rosch00,rosch01,shimshoni03}
\begin{equation}
 M_{ij}\approx \frac{
 i}{\omega L}\prt{\vm{\dot P_i\dot P_j}^{\!R}(\omega) - \vm{\dot
 P_i\dot P_j}^{\!R}(0)},
 \label{mmgen}
\end{equation}
where $\vm{...}^{\!R}$ are the retarded correlation functions
evaluated with respect to the unperturbed Hamiltonian as $\dot
P_i$ is already {\em linear} in the perturbations.
\XXX{Eq.~(\ref{Sigmamatform}) can be rewritten as a coupled rate equation for the
momenta $\vm{P_i}$, in which $\hat\tau^{-1}=\hat M \hat \chi^{-1}$ can
be identified with the matrix of relaxation rates.}

To evaluate (\ref{mmgen}) to leading order, we need various
correlation functions of the decoupled  spin-phonon system. As
discussed above and checked below, for $v_p < v_s$ and low $T$, one
needs high-energy correlation function in the phonon sector but only
the low-energy asymptotics of spin-correlation functions.
To obtain the correct low-energy correlators in the spin-sector it is in
general necessary to take into account the (unitary) scattering of the
thermally excited quasi-particles using generalizations of Sachdev's
semi-classical arguments \cite{sachdev97b} -- see appendix \ref{appGD}. For
example, the correlator $\Cal G_D(x,t)=\vm{\Cal O_D(x,t)\Cal O_D(0,0)}$
of the dimerization operator, which is related to the Majorana fields in a highly
non linear and non local
way, is given by:
\begin{equation}
 \Cal G_D(x,t)\!= \!N\,
 K_0\big( \textstyle\frac{\Delta_0}{v_0}\sqrt{x^2-v_0^2t^2} \big) \; e^{-\frac{1}{2}\Phi\prt{3x/
 \zeta_1,3t/\tau_1}}
 \label{corrOD}
\end{equation}
with $N$ a non universal prefactor, $\Phi(\bar x,\bar t) = \bar x\;\mbox{erf}\left( \bar x/\bar
t\sqrt{\pi}\right) + \bar t e^{-\bar x^2/(\pi\bar t^2)}$, $K_0$ the
modified Bessel function, $\tau_1 =\frac{\pi}{2T}\;e^{\Delta_1/T}$,
and $\zeta_1 = v_1\sqrt{\frac{\pi}{2\Delta_1 T}}\;e^{\Delta_1/T}$. It
will, however, turn out that the scattering from other thermally
excited quasi-particles described by $\Cal G_D(x,t)$ is not important
as our problem is dominated by spin-phonon scattering.

A generic Umklapp memory matrix entry at zero frequency can be cast
in the form $q^2\im\int tdt\int dx e^{iqx} \Cal G_p \Cal G_s$, with
$q=G/2$ or $G$ and $\Cal G_{p(s)}$ the appropriate phonon (spin)
correlator. At low $T<v_p q$, we evaluate it in the saddle point
approximation, by deforming the contour in the complex plane. The
saddle point lies at one of the phonon propagator poles, $v_p
t^*=\mbox{sign}(q)x^*=-\frac{iv_p}{2T}$, well within the  spin
light-cone where there are no contributions from the spin-spin
scattering. Moreover, the semiclassical approximation for $\Cal G_D$
is valid at the saddle point.

The memory matrix can be split into four contributions:
\begin{equation}
 M=M^N+M^{ph}+M^{G/2}+M^{G}
\end{equation}
corresponding to the relaxation processes described by the terms
(\ref{normalsp-ph}), (\ref{pUmk}), (\ref{spUmk1}) and (\ref{spUmk})
respectively.

Generically, a memory matrix entry bears an activated form at low
temperature, and thus can be specified by an activation gap and a
prefactor. While the prefactor depends on non-universal, high-energy
features of the spin-system, in the soft phonon limit $v_p<v_s$ only
\emph{universal} features of the spin system determine the activation
gap. This is why we only give the leading, exponential behavior at low
$T$ for the various memory matrix entries. (with the exception of
  one limiting case, see below). 
These expressions involve
different Umklapp gaps:
\begin{eqnarray}
 E_p &=&\frac{v_p G}{2}, \quad E_D^{}=\frac{v_p
  G}{4}+\frac{\Delta_0}{2}\sqrt{1-\alpha_0^2},\nonumber\\
 E_b&=&\frac{v_p
  G}{2}+\Delta_b\sqrt{1-\alpha_b^2}
 \label{defGapUmk}
\end{eqnarray}
where $\alpha_b=\frac{v_p}{v_b}<1$ and $b=0,1$
labeling the spin singlet and triplet excitations.
These formulas have a simple interpretation: under the constraints of
energy- and momentum conservation they are the lowest energies for
processes where the momentum $q=G$ or $G/2$ is absorbed by scattering a phonon
from $ -q/2 \pm O(\Delta/v_p)$ to $ q/2 \mp O(\Delta/v_p)$ while
scattering ($E_a$) or creating ($E_D$) a spin excitation.
The simple form of the phonon energy scale appearing in Eq.(\ref{defGapUmk}) is a result of our simplified treatment of the phonon sector -- high energy features thereof would modify this form but not the fact that it depends of a single scale $E_p\sim\Theta_D$.

The leading $T$--dependence of the various contributions to the
memory matrix is summarized as follows. The phonon Umklapp
(\ref{pUmk}) gives rise to a single non vanishing entry,
$M^{ph}_{pp}\sim e^{-E_p/T}$. Entries due to normal processes are
$M^N_{pp}=x_0+x_1$, $M^N_{bb}=-M^N_{pb}=x_b$, $x_b\sim
e^{-\Delta_b/T}$.  Spin-phonon Umklapp processes with momentum
transfer $G$ entries are: $M^{G}_{pp}=y_0+y_1$,
$-C_b^{-1}M^{G}_{pb}=C_b^{-2}M^{G}_{bb}=y_b$ with
$C_b=\frac{\alpha_b}{\sqrt{1-\alpha_b^2}}\frac{2\Delta_b}{v_b G}\ll
1$  and $y_b\sim e^{-E_b/T}$. Finally, the relevant $G/2$ entries
are $M^{G/2}_{pp}=-C_0^{-1}M^{G/2}_{p0}=C_0^{-2}M^{G/2}_{00}\sim
e^{-E_D^{}/T}$.
\begin{figure}
\includegraphics[width=\linewidth]{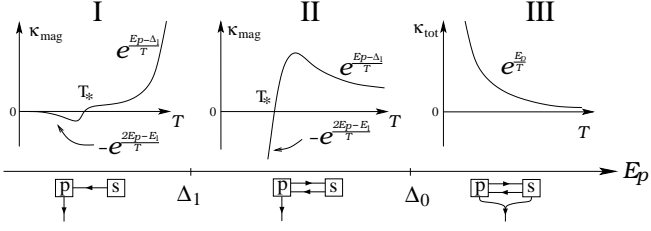}
\caption{Schematic plots of $\kappa(T)$ in three regimes. Depending
on the size of $E_p$ (i.e. the Debye energy) compared to the values
of the spin gaps one obtains qualitatively different behaviors of
the heat conductivity \label{fig1} (shown  for $T \ll \Delta_1$). Note that in regime I and II, $\kappa_{mag}$ is displayed, while in regime III, the total $\kappa$ is displayed, ($\kappa_{mag}$ is not defined in this regime -- see text).
Below the horizontal axis are shown schematically the dominating
momentum transfer processes within the spin and phonon system, and
from those modes via Umklapp scattering to the lattice. $T_* $ is of
order $(v_p/v_1)^2\Delta_1$.} \vspace{-0.5cm}
\end{figure}

With all these elements at hand, using Eq. (\ref{kappasum},\ref{Sigmamatform},\ref{mmgen}) and performing a straightforward matrix inversion yields the heat conductivity, which displays very different behaviors depending on the relative value of the magnetic gaps and Debye energy. In the following, we follow a more physically transparent approach and present in detail the temperature dependence of $\kappa$ as well as the physical mechanism explaining it, in different parameter regimes.

\section{Results and discussion}

For the interpretation of our results detailed below, it is useful
to rewrite the linear-response relation $\vm{P_i}=\sigma_{ij} A_j$
where $A_j$ is an external field coupling linearly to $P_j$. Using
Eq.~(\ref{Je}), we can identify it with $A_j=v_j^2 \frac{\nabla
T_j}{T}$ where $\nabla T_j$ is a fictitious temperature gradient
coupling only to the subsystem $j$. Using Eq.~(\ref{Sigmamatform})
one obtains
\begin{eqnarray}\label{rate}
\frac{\partial}{\partial t} \vm{P_i} - \chi_{i} v_i^2 \frac{
\nabla T_i}{T} = -(\hat{M} \hat{\chi}^{-1})_{ij} \vm{P_j}
\end{eqnarray}
where we used $\chi_{ij}\approx \chi_i \delta_{ij}$.

This equation has a simple interpretation: it is a rate equation for
the momenta and $\tau^{-1}=\hat{M} \hat{\chi}^{-1}$ can therefore be
identified with the matrix of relaxation rates. The matrix of
conductivities can be extracted from the equilibrium solution of the
rate equation.

We will now discuss our results in the three different regimes
depicted in Fig. \ref{fig1}, assuming always $\Delta_1<\Delta_0$ and
$v_s > v_p$. To investigate the relation to experiments, it is
useful to define the magnetic contribution to $\kappa$
\begin{eqnarray}
 \kappa_{mag}=\kappa-\kappa_{ph}^0=\kappa_{ss}+2\kappa_{ps}+
 \delta\kappa_{pp}
 \label{kappamagdef}
\end{eqnarray}
where $\kappa_{ph}^0$ is the conductivity of a hypothetical system
without magnetic degrees of freedom (estimated in experiments from
fits to $\kappa$ perpendicular to the spin ladders). Note that there
is also a negative contribution from the change of the phonon
conductivity $\delta \kappa_{pp}=\kappa_{pp}-\kappa_{ph}^0$ due to
scattering from spin excitations.

\subsection{regime I}

 We first investigate regime I of Fig. \ref{fig1} where $E_p, T\!<\Delta_1$
and momentum relaxation is dominated by phonon Umklapp ($E_p \!<\! E_1,
E_D$). As $\Delta_0>\Delta_1$, we can neglect the singlet mode and
focus on the 2$\times$2 matrix describing the relaxation of
phonon- and triplet momentum to obtain

\vspace{-0.6cm}
\begin{eqnarray}
 \kappa_{pp}&=&v_p^4\frac{\sigma_{pp}}{T}  = \frac{1}{T}\chi_p
 v_p^4\, \tau^U_{p} \mathop{\sim}_{T\ll\Delta_1} e^{E_p/T}
 \label{pp1} \\
 \kappa_{ss} &=&v_1^4\frac{\sigma_{11}}{T}  = \frac{1}{T}\,\chi_1 v_1^4\, \tau^N_{1 \to p} \sim
 e^{-\Delta_1/T}
 \label{kssNormal} \\
 \kappa_{ps} &=&  v_1^2 v_p^2
 \frac{\sigma_{ps}}{T}=\frac{\chi_1 v_1^2}{\chi_p v_p^2}   \kappa_{pp}
 \sim e^{(E_p-\Delta_1)/T}
 \label{kpp-ps}
\end{eqnarray}
where $\chi_i=\chi_{ii}$ (to leading order, $\hat\chi$ is diagonal)
and $1/\tau^U_p=M_{pp}^{ph}/\chi_p \sim e^{- E_p/T}$ is the phonon
Umklapp rate. The "pure spin heat conductivity" $\kappa_{ss}$ --
corresponding to heat being carried by magnons -- is determined by
the exponentially small number of spin excitations, $\chi_1 \sim
e^{-\Delta_1/T}$, and the non-exponential rate of momentum transfer,
$1/\tau^N_{1\to p}=M^N_{11}/\chi_1$ from the spin system to the
phonon system. Most interesting is the drag term $\kappa_{ps}$,
which dominates over $\kappa_{ss}$. As we can neglect momentum
dissipation within the spin-system, even a small coupling of the
spin and phonon systems by normal processes induces 'perfect drag':
both subsystems have the same drift velocity. Therefore the ratio of
heat currents is given by the ratio of energy densities
$\vm{\rho^\e_i}=v_i^2 \chi_i$ \footnote{This relation follows from
$J_\e^i\simeq v_i^2 P_i$ and the energy continuity equation for both
subsystems. For a derivation in the context of charge transport, see
A.~Rosch and N.~Andrei, JLTP \textbf{126}, 1195 (2002).}, and we
find
\begin{equation}
 \kappa_{ps}=\kappa_{pp}\;\vm{\rho^\e_1}/\vm{\rho_p^\e}
\end{equation}
in complete agreement with our memory matrix analysis.

In Eq.~(\ref{pp1}) only the leading behavior of $\kappa_{pp}$ is
shown. To calculate $\delta\kappa_{pp}$ one has to consider the
leading correction arising from  the subdominant mixed spin phonon
Umklapp (\ref{spUmk}) described by the rate
$(\tau^U_{sp})^{-1}=\sum_{ij}M^{G}_{ij}/\chi_p$. We obtain
\begin{equation}
 \delta\kappa_{pp}= -
 \frac{1}{T}\chi_p v_p^4\,\frac{(\tau^U_p)^2}{\tau^U_{sp}} \,\sim
 -e^{(2E_p-E_1)/T}
 \label{deltak}
\end{equation}
This term corresponds to 'magnetic friction': the heat current carried by phonons is reduced due to scattering on the dilute gas of magnons.

The competition between the negative $\delta\kappa_{pp}$ and the
positive $\kappa_{ps}$ leads to a complex cross-over behavior of
the magnetic heat conductivity $\kappa_{mag}$,
see Fig.\ref{fig1}.
Below
$T_* = (2E_p\!-\!E_1)\!-\!(E_p\!-\!\Delta_1)\approx\frac{\alpha_1^2}{2}\Delta_1$
(for $\alpha_1\!\ll\! 1$), the negative $\delta\kappa_{pp}$ dominates.
 For $\Delta_1\!>\!T\!>\!T_*$,
the behavior is fixed by  the pre-exponential factors ; in
this regime we find
\begin{equation}
 \frac{\kappa_{ps}}{|\delta\kappa_{pp}|}= \frac{\Cal N}{\alpha_1}
 \prt{\frac{v_p G}{\Delta_1}}^4\;
 \prt{\frac{T}{\Delta_1}}^{-9/2}\; \prt{\frac{\bar g^U_{ph}}{\bar
 g^U_{sp-ph}}}^2
\end{equation}
where the $\bar g^U$'s are the dimensionless microscopic couplings
and $\Cal N$ is a numerical constant. Interestingly, the cross-over
scale $T_*$ corresponds to the temperature where the saddle point
location crosses the thermal wavelength of the magnons.

Under the hypothesis that for $T>T_*$, magnetic friction is dominated by the enhancement of conductivity due to the addition of heat carriers, namely the magnons, we can relate the ratio of the magnetic versus phononic contributions to the heat conductivity, which is of experimental relevance, to purely thermodynamical quantities:
\begin{equation}
\frac{\kappa_{mag}}{\kappa_{ph}^0} \simeq \frac{v_1^2\chi_{11}^{}}{v_p^2\chi_{pp}^{}}.
\end{equation}

\subsection{regime II}

Next, we consider regime II of Fig.~\ref{fig1}, where $\Delta_1 <
E_p < \Delta_0$. In this regime, the formulas
(\ref{pp1},\ref{kpp-ps},\ref{deltak}) for $\kappa_{pp},
\kappa_{ps}$ and $\delta \kappa_{pp}$ remain unchanged, however,
in contrast to regime I, both $\kappa_{ps}$ and $\delta
\kappa_{pp}$ are exponentially large. Furthermore, the momentum
transfer rate $1/\tau_{p \to 1}$ from the phonon system to the
spin system is now larger than the phonon Umklapp rate
$1/\tau^U_p$, and therefore the two subsystems equilibrate before
loosing momentum via phonon Umklapps, yielding:
\begin{equation}
 \kappa_{ss} = \frac{1}{T}\,\chi_1 v_1^4\, \frac{\tau^N_{1\to p}}{\tau^N_{p\to
 1}}\tau^U_{p}
 \sim e^{(E_p-2\Delta_1)/T}.
 \label{kssUmklapp}
\end{equation}
$\kappa_{ss}$ is always smaller than
$\kappa_{pp}$, and  generically $\kappa_{ss}\ll\kappa_{ps}$.
In this regime, the same remark as in regime I, regarding the relative size of the magnetic versus phononic heat conductivity, applies: if magnetic friction is subdominant, the ratio only depends on thermodynamical quantities.

\subsection{regime III}

In regime III, $E_p > \Delta_0$, a new scattering channel
(\ref{spUmk1}) dominates where (pseudo-) momentum in quanta of
$G/2$ is transferred to the lattice in a complex process involving
singlet, triplet and phonon modes. The associated Umklapp gap
$E_D<E_p$ therefore replaces $E_p$ in formulas
(\ref{pp1},\ref{kpp-ps},\ref{kssUmklapp}). The leading term is
 the phonon contribution
\begin{equation}
 \kappa_{pp}=\frac{1}{T}\,v_p^4\chi_p \tau^U_D\sim e^{E_D^{}/T}
 \label{kppD}
\end{equation}
with $1/\tau^U_D=\sum_{ij}M^{G/2}_{ij}/\chi_p$ the rate of total
momentum relaxation. As the phonon contribution is strongly reduced
by the presence of the magnetic system, the naive procedure of
disentangling the magnon contribution $\kappa_{mag}$ by subtracting
the phonon background [Eq.~(\ref{kappamagdef})] is not useful in
this regime.

\subsection{comparison to experiments}

The experimental data presented in Ref.~[\onlinecite{ladder}] on
cuprate ladders
clearly display an activated behavior for $\kappa_{mag}$,
the activation energy being close to the gap value (which is of the
order of 400K). Measurements have been carried on for temperature
ranging from a few tens of kelvins up to 300K. Due to the operative
way to determine the magnetic heat conductivity, namely by
subtracting a 'pure phonon' contribution obtained by a low
temperature fit, these data are reliable only for not too low
temperatures, i.e. $T\gtrsim50K$. 

Our theory does {\em not} directly apply to these systems as the
  low-temperature properties are dominated by disorder.
Nevertheless, it is possible to obtain a qualitative understanding
  what happens in a situation where the phonon sector is disordered.
Then, due to the scattering from defects (in combination with normal
phonon-phonon processes \cite{ziman63}), the phonon heat current
$J_\e^{ph}$ has no longer an exponentially large lifetime. It is
possible to mimic this situation by taking the limit $E_p\to 0$ in our
equations. This indicates that these compounds are located deep into
regime I, and the corresponding activation energy for $T>T_*$ is then
close to the spin gap. Unfortunately, the temperature $T_*\sim 4K$ turns
  out to be much lower than the minimal temperature above which the
  spin contribution to the heat conductivity can be determined
  reliably in the experiment. Therefore the characteristic sign change
shown in the first panel of Fig.~\ref{fig1} is not observable in these samples.

\section{conclusion}

Our results emphasize the richness of
transport phenomena, the actual low $T$ heat conductivity depending in
a subtle way on the different scales present in the system. In
particular, we find that in a clean system the magnetic contribution
to the conductivity does not display the trivial activated behavior
with activation energy the spin gap. An exception is the limit
$E_p \to 0$ which mimics a situation where momentum predominantly
relaxes via disorder in the phonon system, as might be
the case in the cuprate systems of Ref.[\onlinecite{ladder}]. We hope there will be in the future more measures on spin-ladder materials allowing to probe the different regimes discovered by our study.

The authors would like to thank S. Sachdev and A. Sologubenko for
stimulating discussions, and the German-Israeli Foundation (GIF), the DFG
under SFB 608 and the NSF under DMR 0312495 for financial support.

\appendix

\section{long distance spin correlators}
\label{appGD}

In this appendix, we present the calculation for the low temperature
dimerization operator correlator, $\Cal G_D(x,t)=\vm{\Cal
O_D(x,t)\Cal O_D(0,0)}$, which relies on the semi-classical approach
developed in Ref.[\onlinecite{sachdev97b}]. We first recall the
results obtained by Sachdev and collaborators for a single Ising
model, and we then apply these ideas to our system, which consists
of four weakly coupled Ising models.

\subsection{Single Ising model}

The central idea of this approach is to exploit the fact that at low
enough temperatures in a gapped system, $T\ll\Delta$, typical
configurations of the system correspond to a very dilute gas of
quasiparticles, with mean quasiparticle separation of the order of
$e^{\Delta/T}$, much larger than their thermal de Brooglie wavelength
$\lambda_{th}=\frac{v}{\sqrt{\Delta T}}$. As shown in
Ref.[\onlinecite{sachdev97b}], a classical treatment of these
quasiparticles is legitimate. Only the scattering rates of two
  particles (diluteness allows to consider only two-body collisions)
  has be calculated from quantum mechanics. In one dimension,
  particles with a quadratic dispersion are perfectly reflected from
  any potential in the limit of small momentum. Therefore the two-body scattering matrix
tends to its so-called super-universal form, $S=-1$, irrespective of
the form the two-body potential.  Remarkably, these ingredients are
sufficient to allow for a closed form evaluation of the Ising operator
correlator, both in the ordered and disordered phase, with the result:

\begin{eqnarray}
G_{ord}&=&\vm{\sigma(x,t)\sigma(0,0)} = N^2\;e^{-\Phi(x/\zeta,t/\tau)} \nonumber\\
\Phi(\bar x,\bar t) &=& \bar x\;\mbox{erf}\left( \frac{\bar x}{\bar t\sqrt{\pi}}\right) + \bar t e^{-\bar x^2/(\pi\bar t^2)}
\end{eqnarray}
with $N=\vm{\sigma}$ the local $T=0$ magnetization, $\zeta = v\sqrt{\frac{\pi}{2\Delta T}}\;e^{\Delta/T}$, and
$\tau =\frac{\pi}{2T}\;e^{\Delta/T}$.
In the disordered phase, the correlator reads:
\begin{equation}
G_{dis}=\vm{\sigma(x,t)\sigma(0,0)}=\Cal A \; K_0\left(\textstyle\frac{\Delta}{v}\sqrt{x^2-v^2t^2}\right)  \; e^{-\Phi(x/\zeta,t/\tau)}.
\end{equation}

We are interested in the correlator of the dimerization operator
$\Cal O_D=\prod_{a=0}^3\sigma^a$. The crudest approximation consists
of neglecting the interaction between the Ising models, and  leads
to a simple product form,  
$\Cal G_D=G_{dis}^0\prod_{a>0}G_{ord}^a$.
Of course, such an approximation is expected to fail to capture the
correct long time and space limit, since it neglects the scattering
between quasiparticles on different Ising models. In reality, the
Ising models are coupled -- to lowest order, this coupling is given
by the spin density-spin density coupling $\sum_{a,b} \int
\xi_\R^a\xi_\R^b\xi^a_\L\xi^b_\L$ -- and the scattering is
\emph{relevant} in the sense that it qualitatively affects the form
of $\Cal G_D$ at large $t,x/v\gg \Delta^{-1}$. We now proceed to
take this interaction into account, in the limit of heavy singlet
excitations, $\Delta_0>\Delta_1$. This is a priori the relevant
physical regime, since in actual realizations of the spin ladder no
indication for the existence of the singlet branch, at reasonably
low energy, has ever been reported to our knowledge.

\subsection{Weakly coupled Ising models}

In the semi classical approximation, field configurations
contributing to the path integral representation of the correlation
function are classical ones: quasi particles follow straight lines,
each been given its corresponding Boltzmann weight. Then, the
interaction between two particles is treated quantum mechanically,
the $S$ matrix bearing its super universal form ($k\rightarrow 0$
limit): particles bounce on each other (hardcore collisions) with a
$\pi$ phase shift ($S=-1$). We now repeat Sachdev {\it et al.}'s
\cite{sachdev97b} line of reasoning including inter Ising model interaction; this
amounts to have the straight lines representing the propagation of
thermally excited states to "see" each other, i.e. the
quasiparticles belonging to different Ising models to scatter, the
$S$ matrix being again the super universal one.

For the Ising models $a\neq 0$, that are in the ordered phase, quasi
particles are domain walls for the different Ising models that
separate domains with different magnetization. Each domain wall
carries an index $a\neq 0$. Each time a domain wall $a\neq 0$ crosses
the line joining the points $(0,0)$ and $(x,t)$, the operator
$\prod_{a\neq 0}\sigma^a$ has its $\pm 1$ eigenvalue flipped. This allows us to conclude that its semi
classical correlator reads:
\begin{equation}
\vm{\prod_{a\neq 0}\sigma^a(x,t)\prod_{a\neq 0}\sigma^a(0,0)} = N_1^6 \; e^{-\Phi\prt{3x/\zeta_1,3t/\tau_1}}
\end{equation}
It is the same as the correlator in a single ordered Ising model, the
factor of 3 corresponding to the tripling of the density of excited
states (the probability that a given excited states carrying momentum
$p$ crosses the line joining $(0,0)$ to $(x,t)$ is $q'=3q=3|x-v_{1}(p)
t|/L$, with $L$ the system size and $v_1(p)=\frac{d\epsilon_1(p)}{dp}$
where $\epsilon_1(p)=\sqrt{\Delta_1^2+v_1^2p^2}$ is the one-magnon
dispersion relation).

\begin{figure}[h]
\begin{center}
    \includegraphics[angle=0,width=0.9\linewidth]{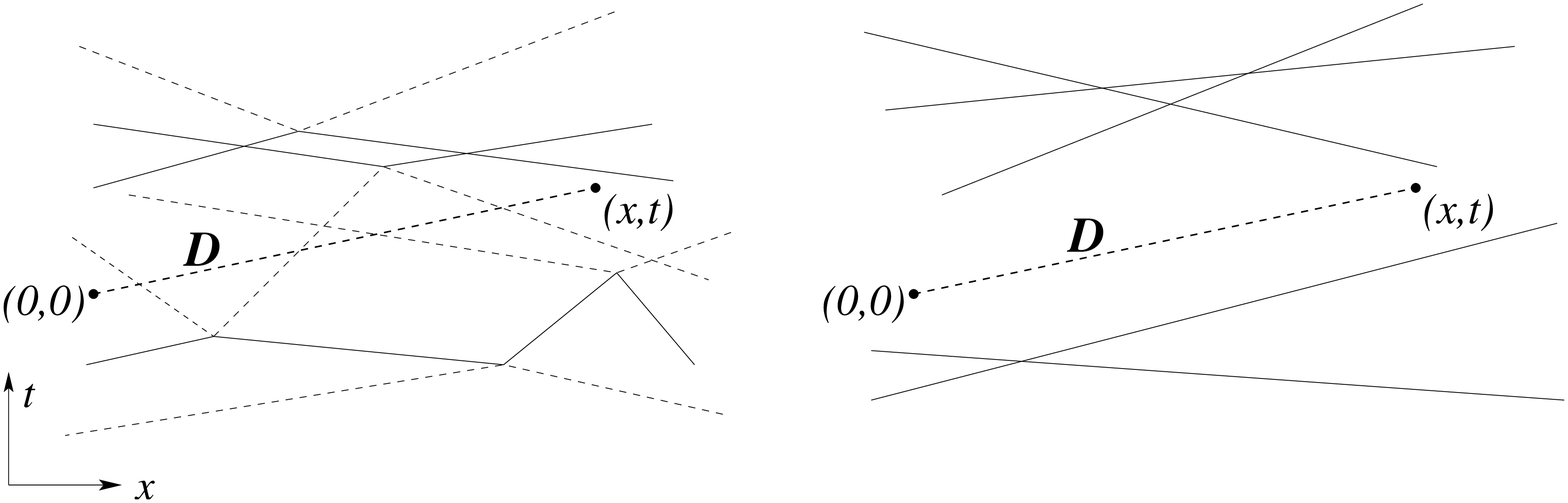}
    \end{center}
  \caption{\label{semiclas}
    Typical configurations contributing to $\Cal G_D$ for different masses. Full lines indicate domain walls of the ordered models $a\neq 0$ while dashed lines represent propagation of quasiparticle of the singlet Ising model. Left: general case. Right: heavy singlet limit.}
\end{figure}

But we are rather interested in the correlator of the dimerization
  operator which includes both singlet and triplet fields. The
singlet Ising model being in its disordered phase, the spin
$\sigma^0\equiv\sigma^{0,z}$ has an overlap with the quasiparticle
creation operator.
Using this remark, the semiclassical calculation can be formulated
  using the real-space diagrams of Sachdev {\it et al.} \cite{sachdev97b}: full lines represent world lines of domain walls in the ordered
models ; dashed lines represent the propagation of excitations of the
disordered model ; there is a line joining $(x,t)$ to $(0,0)$ that
corresponds to the creation of a '0' particle at point $(x,t)$ and to
its destruction at point $(0,0)$ (therefore this line should be dashed
on both its extremities), and this line has to be a straight line
(the state 
$\Cal
O_D(x,t)\Cal O_D(0,0)|\Psi\rangle$ has to have a finite overlap with
$|\Psi\rangle$ to contribute to the trace defining the expectation
value ; therefore the pattern of world-lines with and without the 
straight line connecting $(0,0)$ and  $(x,t)$ has to be the same)
-- in the following we call
this straight line $D$. Additional rules are: each time a dashed line
crosses $D$, there is a scattering event, particles bounce onto each
other and this contributes a $S=-1$ factor -- note that since we are
in one dimension and because the two scattering particles have the 
same dispersion relation, lines coming out of a scattering event are
just continued straight.

To proceed, one has to analyse collisions between dashed and full lines. A simplification occurs in the case of degenerate masses
 $\Delta_0=\Delta_1$. 
Then, energy-momentum conservation during the scattering events ensures that lines are also continued straight (particles just exchange their quantum numbers), and it can be shown that the problem maps onto the calculation of the Pott's spin
 correlator in the 4-state Pott's model in its disordered phase, a problem which has itself been solved
 in the low $T$ regime using the semi-classical approach\cite{rapp06}. 

This fine-tuned degenerate limit is however not relevant to our situation where masses are generically different.
If $\Delta_0\neq \Delta_1$, during a scattering event between a domain wall and a type '0' particle, particles don't just exchange their momenta and lines are deviated after the crossing. In particular, if one considers collisions between domain walls and the line $D$, and remembering that the pattern of world-lines with and without the line $D$ should be the same, one concludes that these events just don't give any contribution to the semi classical correlator -- thus an additional rule
is to forbid scattering between plain lines and $D$. Note that without $D$-full line crossing, the dimerization operator is never flipped by any domain wall.

Given all these rules, one has to sum up contributions over
configurations with no crossing between full lines and $D$. In
general, this is a tough problem. A simplification occurs in the case
where masses are very different, and, for our purpose, when the
singlet mass is much greater than the triplet mass. Then, it is
possible to neglect all contributions with dashed lines in the thermal
background ; the error is of the order of
$\exp(-(\Delta_0-\Delta_1)/T)$. Enforcing that configurations contributing contain no $D$-full line crossing amounts to the replacement of $(1-2q')^N$ by $(1-q')^N$ in Eq.(3) of Ref.[\onlinecite{sachdev97b}]. Performing the average over $p$ we get:
\begin{equation}
\Cal G_D(x,t)= \Cal A\; K_0\prt{\textstyle\frac{\Delta_0}{v_0}\sqrt{x^2-v_0^2t^2}}\; e^{-\frac{1}{2}\Phi\prt{3x/\zeta_1,3t/\tau_1}}
\end{equation}
for $x,t>0$.

\begin{figure}[h]
\begin{center}
    \includegraphics[angle=0,width=0.45\linewidth]{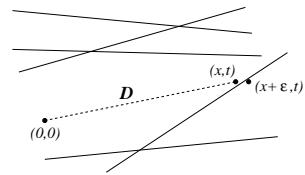}
    \end{center}
  \caption{\label{semiclassder}
    Typical configuration contributing to $\Cal G_D'$ in the heavy singlet mass limit.}
\end{figure}

We will also need to evaluate correlators involving derivatives of just one Ising spin of the kind $\vm{\p_x\sigma^a\prod_{b\neq a}\sigma^b(x,t)\Cal O_D(0,0)}$. Of course, they are not all independent. It suffices for our purpose to compute
\begin{equation}
\Cal G_D'=\vm{\sigma^0\p_x\prod_{a\neq 0}\sigma^a(x,t)\,\Cal O_D(0,0)},
\label{defGdprime}
\end{equation}
which we evaluate using $\Cal G_D'=\lim_{\epsilon\to 0} \Big(\vm{\prod_{a\neq 0}\sigma^a(x+\epsilon)\sigma^0(x,t)\Cal O_D(0,0)} -\Cal G_D(x,t)\Big)/\epsilon$.
Contributions to the first term are those with domain walls passing between $(x\!+\!\epsilon,t)$ and $(x,t)$ but not crossing the line $D$ (see fig. \ref{semiclassder}).

The probability of having one single domain wall with momentum $p$ passing is $q_\epsilon=\epsilon\,\Theta(v_{1}(p) t-x)/L$ (with $\Theta$ the Heaviside function), and after calculation we find
\begin{eqnarray}
\Cal G_D'(x,t)&=&\Cal G_D(x,t)\lambda_1(x,t)
\label{resGdprime}\\
\lambda_1(x,t) &=& -\frac{3}{2}\int_{p_0}^\infty \frac{dp}{\pi}\,e^{-\epsilon_1(p)/T} \nonumber
\end{eqnarray}
where $p_0$ is the root of the equation $\frac{\p\epsilon_1(p)}{\p p}=x/t$. We  note that
$|\lambda_1(x,t)|<\frac{3}{2}\,\zeta_1^{-1}$.

\bibliographystyle{unsrt}

\end{document}